\documentclass[preprint,12pt]{emulateapj}
\usepackage{amstext}
\usepackage{apjfonts}
\usepackage{amsmath}
\pdfoutput=1
\begin{document}
\shorttitle{Gamma-Ray Burst Remnants}

\shortauthors{Ramirez-Ruiz \& MacFadyen}

\title{The Hydrodynamics of Gamma-Ray Burst Remnants}

\author{Enrico Ramirez-Ruiz\altaffilmark{1,2} and Andrew
  I. MacFadyen\altaffilmark{1,3}} \altaffiltext{1}{Institute for
  Advanced Study, Einstein Drive, Princeton, NJ 08540, USA}
  \altaffiltext{2}{Department of Astronomy and Astrophysics,
  University of California, Santa Cruz, CA 95064,
  USA}\altaffiltext{3}{CCPP, Department of Physics, New York
  University, 4 Washington Place, New York, NY 10003, USA}

\begin{abstract}
This paper reports on the results of a numerical investigation
designed to address how the initially anisotropic appearance of a GRB
remnant is modified by the character of the circumburst medium and by
the possible presence of an accompanying supernova (SN). Axisymmetric
hydrodynamical calculations of light, impulsive jets propagating in
both uniform and inhomogeneous external media are presented, which
show that the resulting dynamics of their remnants since the onset of
the non-relativistic phase is different from the standard self-similar
solutions.  Because massive star progenitors are expected to have
their close-in surroundings modified by the progenitor winds, we
consider both free winds and shocked winds as possible external media
for GRB remnant evolution. Abundant confirmation is provided here of
the important notion that the morphology and visibility of GRB
remnants are determined largely by their circumstellar
environments. For this reason, their detectability is highly biased in
favor of those with massive star progenitors; although, in this class
of models, the beamed component may be difficult to identify because
the GRB ejecta is eventually swept up by the accompanying SN. The
number density of asymmetric GRB remnants in the local Universe could
be, however, far larger if they expand in a tenuous interstellar
medium, as expected for some short GRB progenitor models. In these
sources, the late size of the observable, asymmetric remnant could
extend over a wide, possibly resolvable angle and may be easier to
constrain directly.
\end{abstract}


\section{Introduction}\label{int}
Relativistic jets are common in the astrophysical environment. Objects
known or suspected to produce them include radio galaxies and quasars
\citep{b84}, microquasars \citep{mr99} and gamma-ray bursts
\citep{g06}. An important difference between jets of gamma-ray bursts
(GRBs) and the better studied radio jets of quasars or microquasars is
that active quasars often inject energy over extended periods of time
into the jet while GRB sources are impulsive.  Although quasar jets
remain highly collimated throughout their lifetimes, GRB jets
decelerate and expand significantly once they become
nonrelativistic. Expansion into a uniform medium has been well studied
\citep{ap01}, but the interaction of a GRB remnant with a nonuniform
medium remains poorly understood.

Much of our effort in this paper is therefore dedicated to determining
how the morphology and dynamics of young GRB remnants is modified by
the character of the circumburst medium.  Some of the questions at the
forefront of attention include the effects of the external medium and
the degree to which GRB remnant dynamics and structures are modified
by the presence of an accompanying supernova. We address both of these
issues here. Because massive stars are expected to have their close-in
surroundings modified by the progenitor winds, we consider both free
winds and shocked winds as possible surrounding media for the GRB
remnant evolution. Detailed hydrodynamic simulations of this
interaction are presented in \S\S \ref{wind} and \ref{bubble}, while a
brief description of the numerical methods and the initial models is
giving in \S \ref{nm}. For completeness, the interaction with a
constant-density medium is discussed in \S \ref{const}. The role of
supernova explosions in shaping the evolution and morphology of GRB
remnants is discussed in \S \ref{sn}.  The effects of a nonspherical
circumburst medium are briefly addressed in \S \ref{wind}. Discussion
and conclusions are presented in \S \ref{dis}.

\section{Numerical Methods and Initial Model}\label{nm}
\subsection{The Underlying Dynamics}\label{ud}
For simplicity, lets consider a uniform GRB jet with sharp edges and a
half-opening angle $\theta_{\rm \theta}$, with an initial value of
$\theta_0$.  The typical angular size, $R_\perp$, of the jet at $t <
t_{\theta}$, where $t_{\theta}$ is the time at which the jet's Lorentz
factor $\gamma$ drops to $\theta_0^{-1}$, is the same as for a
spherical flow:
\begin{equation}
R_\perp (t) \propto \left({E_{\rm iso} \over A}\right)^{1 \over
  2(4-k)}t^{{5-k \over 2(4-k)}}\propto \left({E \over A}\right)^{{1\over 2(3-k)}}t_{\rm
  j}^{{-1\over2(4-k)}}t^{{5-k \over 2(4-k)}}
\end{equation}
for an external density profile $\rho_{\rm ext}=Ar^{-k}$
\citep{grrl05}. Here $E$ is the {\it true} kinetic energy content of
the jet, and $E_{\rm iso}=f_b^{-1}E$ is the isotropic equivalent
energy where $f_b\approx\theta_0^2/2$ is the beaming factor. At these
early times, the flow is described by the Blandford-McKee (1976)
self-similar solution, which provides an accurate expression for the
temporal evolution of the observed image size:
\begin{equation}
R_\perp (t) = 4\times 10^{16}\left({E_{\rm iso,52}\over n_0}\right)^{1/8}\left({t_{\rm
    days} \over 1+z}\right)^{5/8}\;{\rm cm}
\end{equation}
for constant density medium ($k=0$) with $n=1\;n_0$ cm$^{-3}$
($\rho=1.67\times10^{-24}\rho_0$ g cm$^{-3}$), and
\begin{equation}
R_\perp (t) =2.5\times 10^{16}\left({E_{\rm iso,52} \over A_\ast}\right)^{1/4}\left({t_{\rm days} \over 1+z}\right)^{3/4}\;{\rm cm}
\end{equation}
for a stellar wind environment ($k=2$) with $A_\ast = A/5 \times
10^{11}$ g cm$^{-1}$.

Notice that even in the most optimistic cases, the characteristic size
of a GRB image is only of order 1 $\mu$ as about a day after the GRB
at the Hubble distance \citep{gl01}, and so it cannot be resolved by
existing telescopes.  Obviously, the challenge is made easier for
nearby sources, where the late size of the observable remnant could
extend over a wide, possibly resolvable angle
\citep{onp04,grr04,grrl05}. At late times $t > t_{\rm NR}$, where
$t_{\rm NR}$ is the nonrelativistic transition time, the jet is
expected to gradually approach the Sedov-Taylor self-similar solution,
asymptotically reaching $R_\perp (t) \propto (E/A)^{1/(5-k)}
t^{2/(5-k)}$. At $t_{\theta} < t < t_{\rm NR}$ there is, however, a
large uncertainty in the hydrodynamic evolution of the jet, and in
particular in its rate of sideways expansion.

To illustrate the importance of sideways expansion on the evolution of
the observed image, lets consider two extreme assumptions: (i)
relativistic lateral expansion in the comoving frame \citep{r99,s99}
for which $\theta_{\theta}\approx\max(\theta_0,\gamma^{-1})$ so that
at $t_{\theta}<t<t_{\rm NR}$ we have
$\gamma\approx\theta_{\theta}^{-1}\approx\theta_0^{-1}\exp(-R/R_{\theta})$,
and (ii) little or no lateral expansion,
$\theta_{\theta}\approx\theta_0$ for $t<t_{\rm NR}$, in which case
appreciable lateral expansion occurs only when the jet becomes
sub-relativistic and gradually approaches spherical symmetry.

For relativistic lateral expansion, $\theta_{\theta}\sim 1$ at $t_{\rm
  NR}=t_{\rm NR}(E)$, where $R_{\rm NR}(E)=ct_{\rm NR}(E)=[(3-k)E/4\pi
  Ac^2]^{1/(3-k)}$ and the jet radius will be similar to that of the
  Sedov-Taylor solution, $R_{\rm ST}(E,t) \sim (Et^2/A)^{1/(5-k)}$. In
  this case, one expects the flow to approach spherical symmetry only
  after a few dynamical times.  This is probably not the case as
  clearly illustrated by the results of numerical simulations
  \citep{g01,c04,gr07} showing only modest lateral expansion as long
  as the jet is relativistic. If lateral expansion is neglected, the
  jet becomes sub-relativistic only at
\begin{equation}
R_{\rm NR}(E_{\rm iso})=ct_{\rm NR}(E_{\rm iso})=\left[{(3-k)E_{\rm
      iso} \over 4\pi Ac^2}\right]^{1/(3-k)}.
\label{nr}
\end{equation}
For expansion in a constant density medium ($k=0$), eq. (\ref{nr}) can
be rewritten as
\begin{equation}
R_{\rm NR}(E_{\rm iso})=0.3 \left({E_{51}\over \rho_0}\right)^{1/3}
 \left({f_b^{-1} \over 30}\right)^{1/3}\;{\rm pc},
\end{equation}
while for a wind medium ($k=2$), it becomes
\begin{equation}
R_{\rm NR}(E_{\rm iso})= 2 \left({E_{51}\over A_\ast}\right)
\left({f_b^{-1} \over 30}\right)\;{\rm pc}.
\end{equation}

From eq. (\ref{nr}) it follows that $R_{\rm NR}(E_{\rm iso})$ is a
factor of $\sim (E_{\rm iso}/E)^{1/(3-k)}=f_b^{-1/(3-k)}\sim
\theta_0^{-2/(3-k)}$ larger than $R_{\rm NR}(E)=ct_{\rm NR}(E)$ and a
factor of $\sim f_b^{-1/(5-k)}\sim \theta_0^{-2/(5-k)}$ larger than
$R_{\rm ST}[E,t_{\rm NR}(E_{\rm iso})]$. Thus eq. (\ref{nr}) simply
states that the jet keeps its original opening angle,
$\theta_{\theta}\approx\theta_0$ until $t_{\rm NR}(E_{\rm iso})$, and
hence at this time the jet is still far from being spherical. Thus,
once the jet becomes sub-relativistic, we expect it to expand sideways
significantly, and become roughly spherical only when it has increased
its radius by a factor of $\psi$. This should occur roughly at a time
$t_{\rm sph}$ when $R_{\rm ST}(E,t_{\rm sph})=\psi R_{\rm NR}(E_{\rm
iso})$:
\begin{equation}
{t_{\rm sph} \over t_{\rm NR}}(E_{\rm iso})\approx
f_b^{-1/2}\psi^{(5-k)/2}\approx\sqrt{2}\,\theta_0^{-1}\psi^{(5-k)/2}.
\end{equation}
This is a factor of $\sim f_b^{-1/2}\approx 14(\theta_0/0.1)^{-1}$
larger than the expected transition time for relativistic lateral
expansion in the comoving frame. In the sections that follows we
present quantitative estimates of $t_{\rm sph}$ for GRB jets expanding
in variety of circumburst environments.

\begin{figure}
\begin{center}
\includegraphics[scale=0.7]{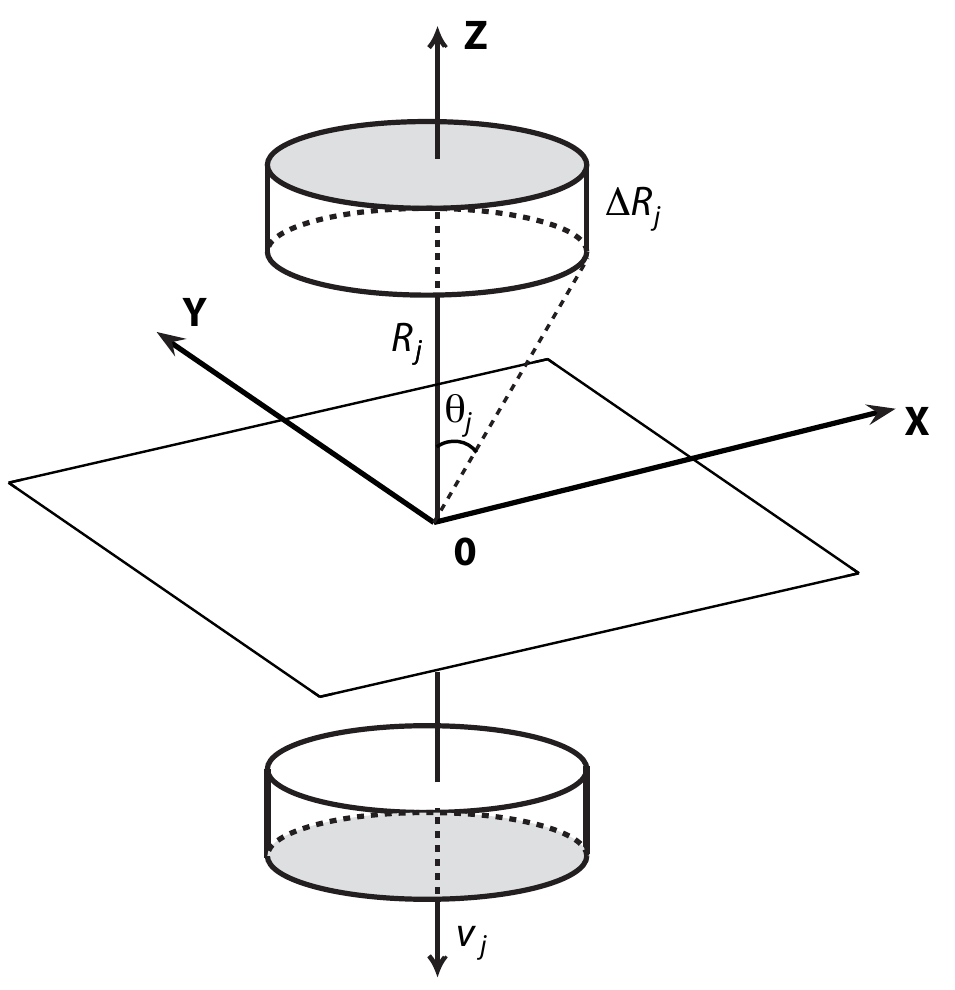}
\caption{Schematic plot illustrating the numerical initiation of the
GRB explosion.}
\label{fig1}
\end{center}
\end{figure}

\subsection{Initial GRB Model}
Common to all calculations is the initiation of the GRB explosion as
two identical blobs expanding in opposite directions into the
circumburst medium. Calculations were done in two dimensions in
cylindrical geometry using the PPM adaptive mesh refinement code FLASH
(ver 2.4). Both blobs and the circumburst medium are modeled by a
cold, $\gamma = 5/3$, ideal gas. The initial configuration is as
follows. The computational domain, as illustrated in Fig. \ref{fig1},
is a unprolonged cylinder in which the ejecta move along the symmetry
axis. In the inner region of each of the pancakes, the ejecta mass,
$M_{\rm j}$ is distributed uniformly. In all runs $v_{\rm j}=0.3c$,
$\Delta R_{\rm j}/R_{\rm j}\approx 0.5$, $\theta_{\rm j} \approx 0.5$,
$M_{\rm j} \approx 2E_{\rm j}/v_{\rm j}^2$ and $R_{\rm j}\approx
R_{\rm NR}(E_{\rm iso})$.

Without a detailed understanding of the exact shape and energy
distribution of the ejecta, we have only an approximate description of
how to construct the initial conditions. However, as clearly
illustrated by \citet{ap01}, the late time evolution of the ejecta is
rather insensitive to uncertainties in the initial conditions. We have
considered various initial densities, angular widths, and shapes of
the collimated ejecta and found that these are indeed unimportant in
determining the late morphology of the remnant. This stems from the
fact that at late times the mass of the remnant is dominated by the
circumburst gas, which washes out any variations in the initial
conditions of the ejecta.

\section{The appearance of a GRB Remnant in the ISM}\label{const}
In this section we present a quantitative discussion of how the GRB
remnant morphology is modified by expansion into a uniform
medium. Expansion into a constant density medium is expected in a
variety of progenitor models, in particular those related to short
GRBs \citep{f99,bl99,b06,lrr07}. The discussion largely follows that
of \citet{ap01}, although analytical solutions are derived here to
illustrate what may not be obvious from earlier derivations.

In the absence of characteristic scales in stellar ejecta and in the
ambient medium, self-similar, spherically symmetric solutions exist,
and they are widely used to interpret observational data of GRB
remnants \citep[e.g.][]{w98}. However, as argued in \S \ref{ud}, near
the non-relativistic transition the remnant is far from being
spherical and thus the Sedov-Taylor solution fails to provide an
accurate description of the evolution of the GRB ejecta. A simple
estimate for $t_{\rm sph}$ can be, however, obtain by assuming that
the expansion of each of the collimated blobs is accurately described
by a self-similar, spherically symmetric solution. Under these
conditions, the problem reduces to depositing a finite amount of
energy $E$ at two different locations separated by a distance $R_{\rm
j}\sim R_{\rm NR}(E_{\rm iso})$. This is valid as long as the external
medium is uniform around the explosion site. The evolution of the
shock radius for each of the blobs will then follow:
\begin{equation}
R_{\rm ST}(t)=\xi \left({E \over \rho_0}\right)^{1/5}\;t^{2/5}
\end{equation}
with $\xi=1.17$ for $\gamma=5/3$. The ratio between the remnant width
and height,
\begin{equation}
\varsigma={R_{\rm ST}(t)\over R_{\rm ST}(t) + R_{\rm j}},
\end{equation}
will approach unity as the two blobs expand and merge. The GRB remnant
will then become nearly spherical in shape after a time
\begin{equation}
t_{\rm sph} \approx 243 \left({E_{51} \over
\rho_0}\right)^{1/3}\left({f_b^{-1} \over
30}\right)^{5/6}\left({\varsigma_{0.9} \over 1-
\varsigma_{0.9}}\right)^{5/2}\;{\rm yr},
\label{sedov}
\end{equation}
when $\varsigma=0.9\varsigma_{0.9}$. Obviously, the above calculation
is only sketchy and should be taken as an order of magnitude figure,
as it not only assumes that the mass of the collimated ejecta is
negligible with respect to the swept-up gas but also neglects the
presence of shocked material throughout the interaction region.

\begin{figure}
\begin{center}
\includegraphics[scale=0.55]{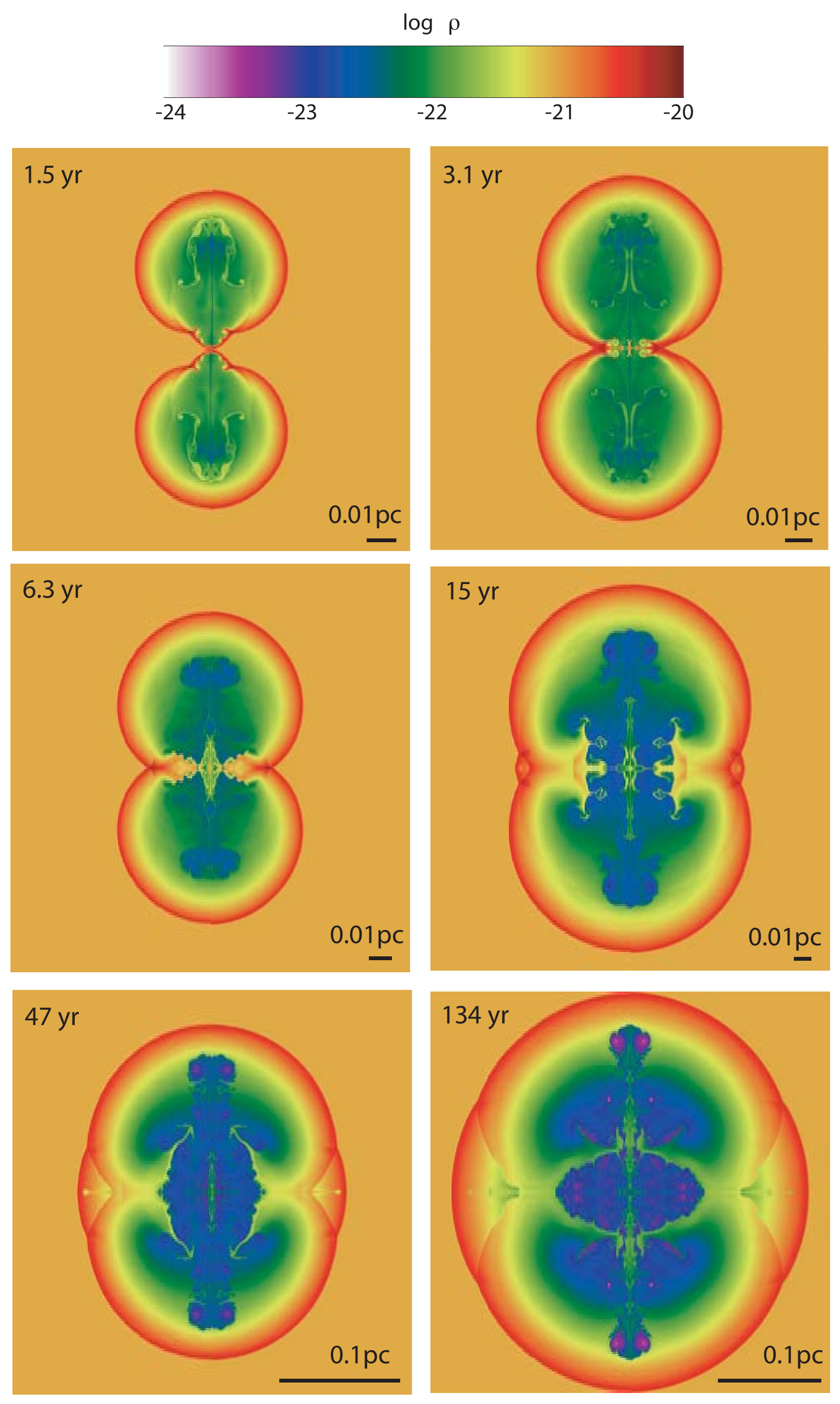}
\caption{The evolution of a GRB remnant in a constant density
medium. The ejecta, $E_{\rm j}=10^{50}$ erg, and surrounding ISM,
$\rho_0=10^3$, are characterized by a 5/3 adiabatic index. Shown are
logarithmic density cuts in g cm$^{-3}$. Calculations were done in
two-dimensional cylindrical coordinates for seven levels of
refinement. The size of the computational domain was (0.3 pc)$^{2}$. }
\label{fig2}
\end{center}
\end{figure}

\begin{figure}
\begin{center}
\includegraphics[scale=0.45]{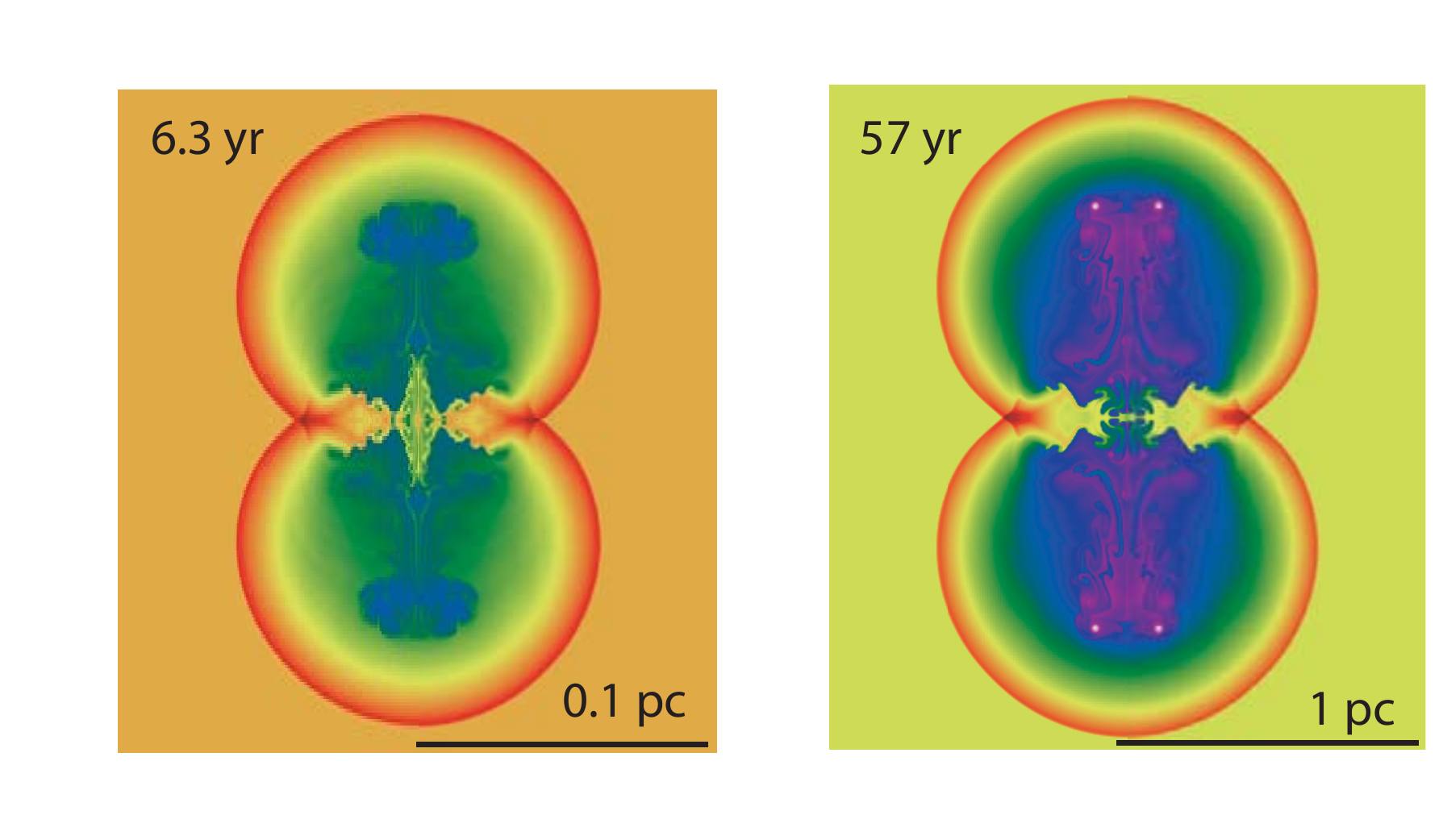}
\caption{The evolution of a GRB remnant for $\rho_0=10^3$ ({\it left})
and $\rho_0=1$. Evolutionary ages in years are indicated in each frame
together with corresponding size scales. As expected for a constant
density medium, a unique combination of $E$, $\rho$ and $t$ has the
dimensions of $R$.}
\label{fig3}
\end{center}
\end{figure}

Detailed hydrodynamic simulations of the evolution of a GRB remnant in
a uniform medium are presented in Fig. \ref{fig2}, where the density
contours of the expanding collimated ejecta at various times in its
hydrodynamical evolution are plotted. As the collimated ejecta
collides with the ambient medium, a bow shock forms. The shock
propagates in the direction of motion but also perpendicular to it
and, over time, wraps around the expanding ejecta. Although initially
the remnant may be highly nonspherical, the ratio between its width
and height will approach unity as the two blobs expand, merge into a
single structure and then finally become spherical in shape. The
resultant structure will not be perfectly spherical as some density
inhomogeneities around the equator resulting from the encounter remain
visible. It will be, however, difficult to distinguish it (based on
morphology alone) from a supernova remnant after about
\begin{equation}
t_{\rm sph} \approx 3 \times 10^3\left({E_{51}\over
\rho_0}\right)^{1/3}\;{\rm yr},
\end{equation}
when $\varsigma\sim0.9$. The governing parameters of the late
evolution of the remnant are the initial energy of the jet, $E$, and
the density of the ambient medium, $\rho$ \citep{ap01}.  As
illustrated in Fig. \ref{fig3}, these two initial parameters can also
determine the early evolution of the remnant for a fixed $f_b$.

\section{Evolution in  a Circumstellar Wind Medium}\label{wind}
If the progenitors of GRBs are massive stars then there is an analogy
to the explosions of core collapse supernovae, for which there is
abundant evidence that they interact with the winds from the
progenitor stars. In most supernova cases, the radial range that is
observed is only out to a few pc, such that the mass loss
characteristics have not changed significantly during the time that
mass is supplied to the wind \citep{cl00}. The density in the wind
depends on the type of progenitor. Red supergiant stars, which are
thought to be the progenitors of Type II supernovae, have slow dense
winds. Wolf-Rayet stars, which are believed to be the progenitors of
Type Ib/c supernovae and possibly of long GRBs \citep{w93,mw99}, have
faster, lower-density winds. The winds from WRs are characterized by
mass-loss rates $\dot M\approx 10^{-5}\,M_\odot\;{\rm yr}^{-1}$ and
velocities $v_w\approx 10^3\,{\rm km\;s^{-1}}$ \citep{cm86}. In a
steady, spherically symmetric wind, the stellar density is $\rho_{\rm
ext}=Ar^{-2}$, where $A=5\times 10^{11}(\dot{M}/10^{-5}M_\odot\;{\rm
yr}^{-1})(v_w/10^3\;{\rm km \; s^{-1}})^{-1}\;{\rm g\;cm^{-3}}$. Note
that for this choice of stellar wind parameters $A_\ast=1$.
\begin{figure}
\begin{center}
\includegraphics[scale=0.45]{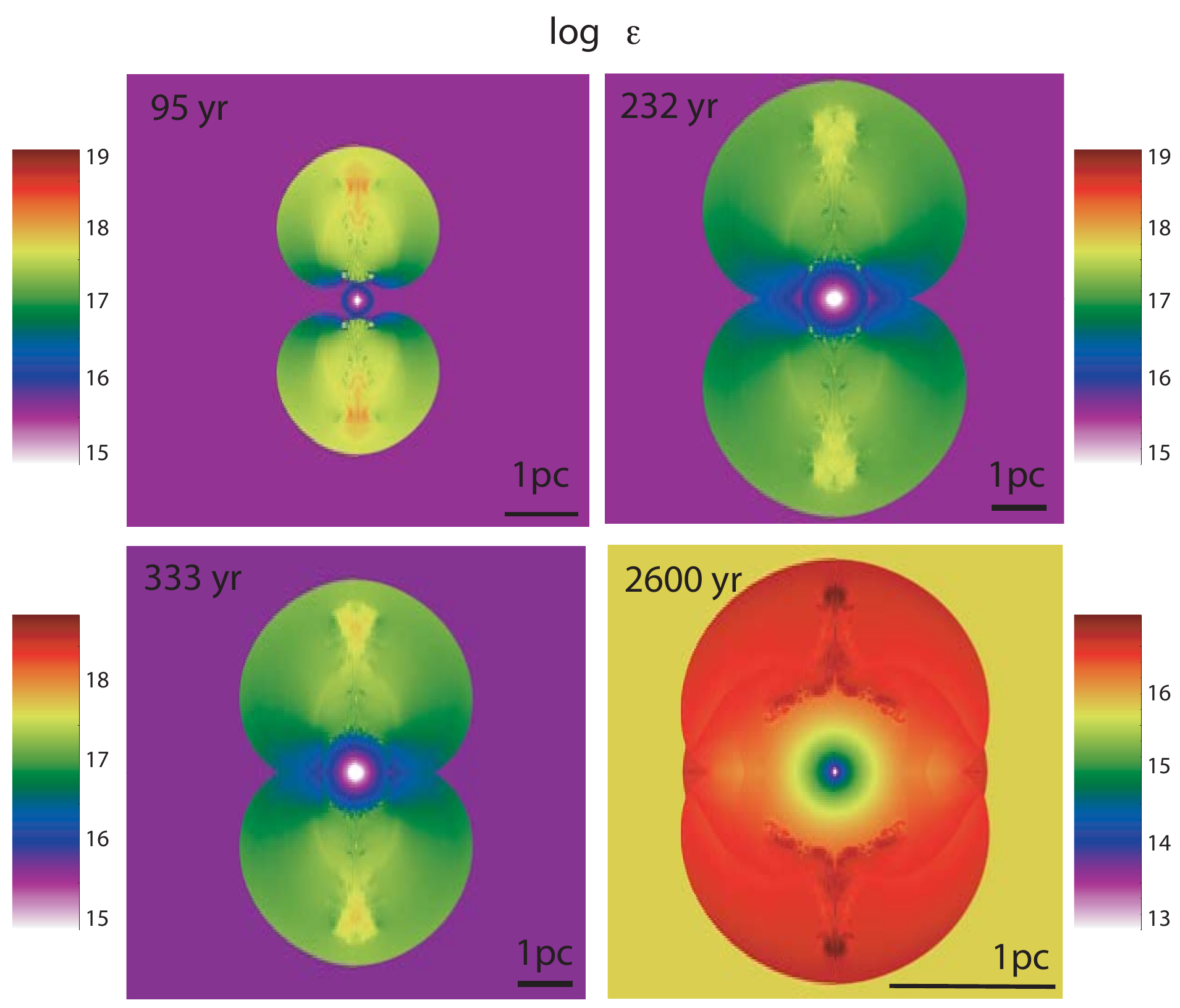}
\caption{The evolution of a GRB remnant in a $1/r^2$ medium.  The
 ejecta, $E_{\rm j}=10^{51}$ erg, and surrounding stellar wind medium
 ($v_w=10^3\,{\rm km\;s^{-1}}$ and $\dot M=2.5 \times
 10^{-5}\,M_\odot\;{\rm yr}^{-1}$) are characterized by a 5/3
 adiabatic index. Shown is the evolution of the specific energy,
 $\epsilon$, in erg/g. Calculations were done in two-dimensional
 cylindrical coordinates for ten levels of refinement. The size of the
 computational domain was (16 pc)$^{2}$.}
\label{fig4}
\end{center}
\end{figure}

For this discussion we shall first assume the stellar wind is
effectively spherical. The evolution of a GRB remnant in a stellar
wind since the onset of the non-relativistic phase is summarized in
Fig. \ref{fig4}. Similar resulting structures to those described in \S
\ref{const} are clearly seen. A bow shock forms as each blob collides
with the stellar wind, which eventually wraps around the ejecta before
the two expanding shells collide to form a single structure. However,
because in a wind medium the swept-up mass increases only linearly
with radius, the GRB remnant decelerates much more slowly than in a
uniform medium. Moreover, in a wind medium, resistance to sideways
expansion is increased. This is because the bow shock, as it wraps
around the ejecta, encounters a steadily increasing ambient
pressure. As a result, the remnant will become roughly spherical only
after a time
\begin{equation}
t_{\rm sph} \approx 5\times 10^4\;\left({E_{51}\over
A_\ast}\right)\;{\rm yr},
\label{tsph}
\end{equation}
when $\varsigma \sim 0.9$.  Beyond this point, the evolution will
evolve into a classical Sedov-Taylor supernova remnant evolution.

The estimate given by equation (\ref{tsph}) could be inaccurate for a
number of reasons. Depending upon the wind history of the progenitor
star and the properties of the surrounding ISM, the density structure
around $R_{\rm sph}\approx 30$pc could be quite complicated. The
non-steady nature of the winds in massive stars together with the
relatively large ISM pressure expected in star-forming regions, leaves
open the possibility of interaction with denser material at much early
times. In this case, the GRB remnant will start being decelerated by
the external medium at a smaller radius than it would expanding into a
free $1/r^2$ wind. Much of our effort in \S \ref{bubble} will
therefore be dedicated to determining the contribution of the
presupernova ejecta of Wolf-Rayet stars to the circumburst
environment, and describing how this external matter can affect the
observable characteristics of GRB remnants.

\begin{figure}
\begin{center}
\includegraphics[scale=0.45]{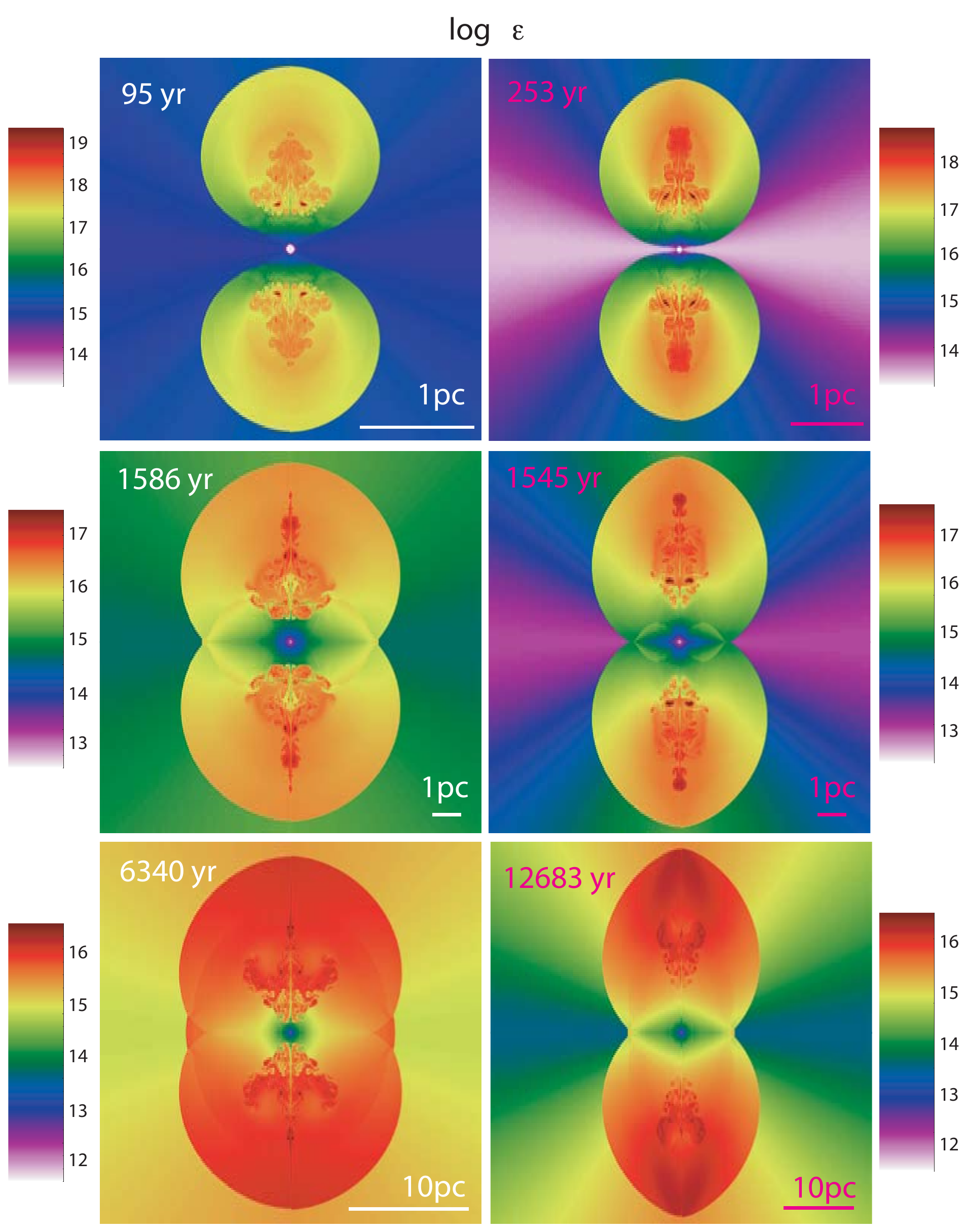}
\caption{The evolution of a GRB remnant in a non-spherical $1/r^2$
 medium.  The ejecta, $E_{\rm j}=10^{51}$ erg, and surrounding stellar
 wind medium ($v_{\rm esc}=10^3\,{\rm km\;s^{-1}}$, $\zeta=1$
 ,$\varphi=0.5$, $\dot M=2.5 \times 10^{-5}\,M_\odot\;{\rm yr}^{-1}$,
 and $\Omega=0.8$) are characterized by a 5/3 adiabatic index. Shown
 is the evolution of the specific energy, $\epsilon$, in
 erg/g. Calculations were done in two-dimensional cylindrical
 coordinates for seven levels of refinement. The size of the
 computational domain was (30 pc)$^{2}$.}
\label{fig5}
\end{center}
\end{figure}

Large-scale density gradients in the ambient medium could result in
asymmetric, nonradial distortions. In this case, $t_{\rm sph}$ could
be larger than that given by equation (\ref{tsph}). For example,
Fig. \ref{fig5} shows the evolution of a GRB remnant in an asymmetric
wind, under the assumption that the progenitor star experiences
non-spherical mass loss close to critical rotation: in other words, a
scenario in which a slower and denser wind is confined to the
equatorial plane. To compute the latitudinal dependence of the wind
properties of a star close to critical rotation ideally requires
multi-dimensional models of the star and its outflowing atmosphere,
which are not available.  \citet{lan98}, however, argued that the
stellar flux and the radius might still vary only weakly from pole to
the equator in very luminous stars. We therefore applied equations
similar to those found by \citet{bc93} for winds of rotating stars in
the limit of large distance from the star:
\begin{equation}
v_{\infty}(\theta) = \zeta v_{\rm esc} \left(1 - \Omega\, \sin\theta
\right)^{\varphi}\;,
\end{equation}
where we set the parameters defined in \citet{bc93} to $\zeta=1$
,$\varphi=0.5$, $\Omega=v_{\rm rot}/ v_{\rm crit}=0.8$, and $v_{\rm
  crit}=v_{\rm esc}/\sqrt{2}=[GM_*(1-\kappa)R_*)]^{1/2}$, with $M_*$
and $R_*$ being mass and radius of the star, and $\kappa$ standing for
the ratio $L/ L_{\rm Edd}$ of stellar to Eddington luminosity. Under
the above conditions, the GRB remnant expands more quickly and easily
into the lower density wind at the poles, producing an increasingly
asymmetric double-lobed structure.

Finally, the estimate given by equation (\ref{tsph}) would be modified
if the beamed GRB is accompanied by an underlying supernova, as
expected in the collapsar model \citep{mw99,mwh01,wb06}. The
large-angle SN outflow, responsible for exploding the star and
producing the ${}^{56}$Ni, would generally carry more energy and
inertia than the relativistic jet itself
\citep[e.g.,][]{s04,m06,kaneko}, so that the latter always overtakes
it, and sweeps up the GRB ejecta. It is to this problem that we now
turn our attention.

\section{Interaction with an Underlying Supernovae}\label{sn}

It seems likely that GRBs originate in a very small fraction of
massive stars that undergo a catastrophic energy release event toward
the end of their evolution. Expressly, the association of some GRBs
with type Ic supernovae \citep[e.g.,][]{hjorth,snspec,elena} has
pointed a finger at deaths of massive Wolf-Rayet stars as the cause of
GRBs, or at least a subset thereof. The central engine is believed to
give rise to a polar outflow with two components
\citep{mw99,rm02,wb06}. One large-angle outflow (the SN), containing
most of the energy and mass, is responsible for exploding the star and
producing the ${}^{56}$Ni to make the SN bright.  A second outflow
component (the GRB jet) occupies a narrower solid angle and probably
contains less energy (which can range from comparable to much less),
and most of its energy is in material with relativistic velocities
(where the typical Lorentz factor of the material that carries most of
the energy in this component can vary significantly among SN-GRBs).

The large-angle SN outflow, carrying more energy and inertia than the
relativistic jet itself, will generally sweep up the GRB ejecta before
it has been much decelerated. An order of magnitude estimate for
$t_{\rm sph}$ can be obtain by assuming that the dynamics of the
laterally-expanding, GRB ejecta is accurately described by a
self-similar, spherically solution in a $1/r^2$ medium and that the SN
outflow does not appreciably slow down. The evolution of the shock
radius of the beamed remnant is then given by
\begin{equation}
R_{\rm ST}(t)=\xi_w \left({E \over A}\right)^{1/3}\;t^{1/3}
\end{equation}
with $\xi_w=0.73$ for $\gamma=5/3$. The GRB ejecta, as it clears the
surrounding stellar matter, will be overtaking by the large
scale-outflow SN at
\begin{equation}
t_{\rm sph} \approx 910 \left({E_{51} \over A_\ast }\right)
\left({\beta_{\rm SN} \over 0.1}\right)^{-3}\;{\rm yr},
\label{sngrb}
\end{equation}
where $\beta=v_{\rm SN}/c \ll 1$. After this time, the merged system
will quickly become spherical. Yet it is clear that this simple
estimate is inadequate as a model for the real complex dynamics, which
necessitates the use of hydrodynamical calculations.

\begin{figure}
\begin{center}
\includegraphics[scale=0.52]{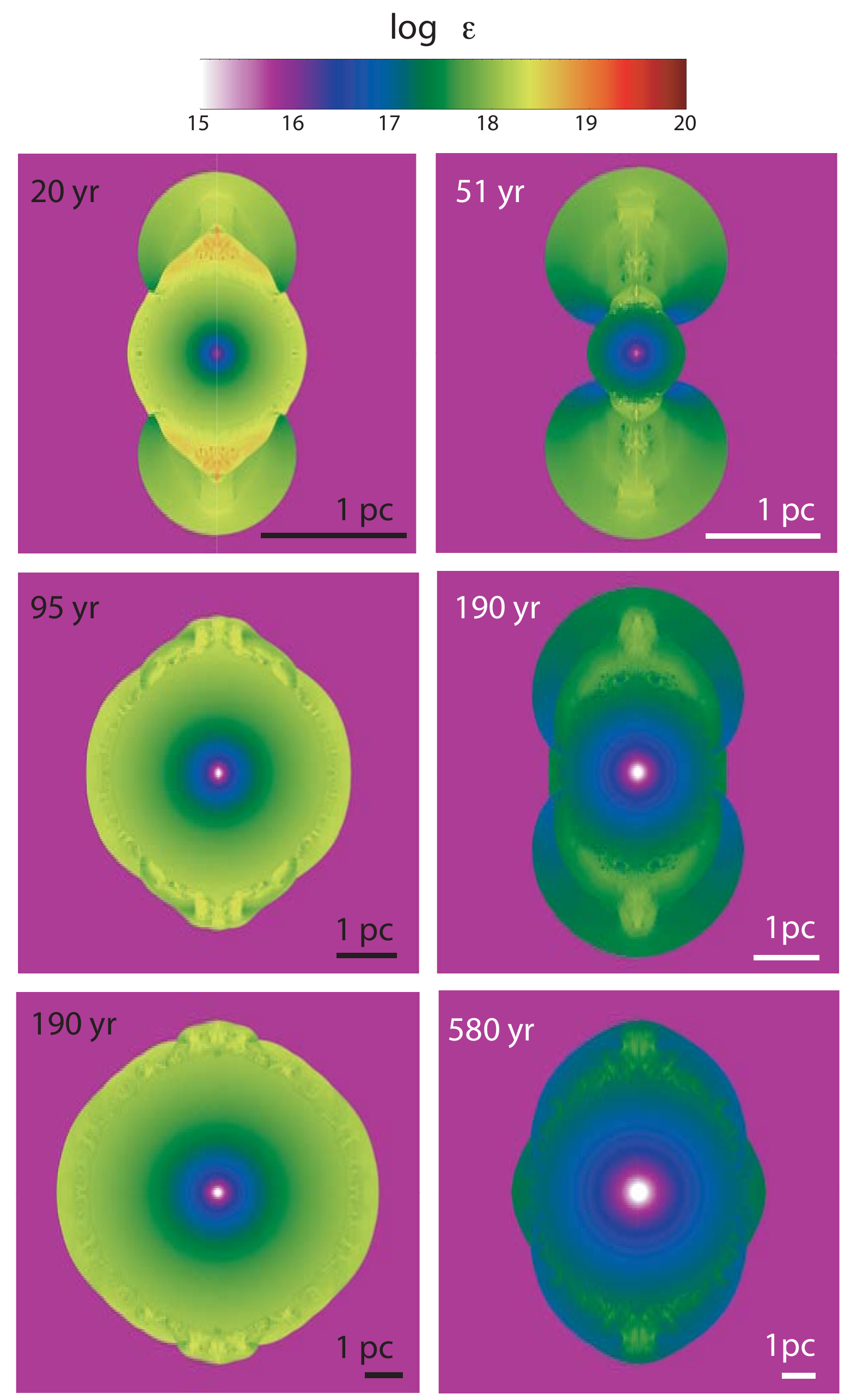}
\caption{The evolution of a GRB remnant interacting with an underlying
(slower expanding) spherical supernova. The stellar wind parameters as
well as the GRB ejecta initial quantities are the same as in
Fig. \ref{fig4}. Shown is the evolution of the specific energy,
$\epsilon$, in erg/g for two different SN explosion energies: $5
\times 10^{51}$ erg ({\it left}) and $5 \times 10^{50}$ erg ({\it
right}). Calculations were done in two-dimensional cylindrical
coordinates for ten levels of refinement. The size of the
computational domain was (30 pc)$^{2}$.}
\label{fig6}
\end{center}
\end{figure}

The evolution of a GRB remnant accompanied by an underlying supernova
is shown in Fig. \ref{fig6} for two different SN explosion
energies. The stellar wind parameters are chosen so that they are
equal to those displayed in Fig. \ref{fig4}. Compared to
Fig. \ref{fig4}, significant structural differences appear when the SN
outflow has grown significantly in size and it starts to overtake the
laterally expanding GRB ejecta. Two illustrative cases are depicted:
$E_{\rm SN}$: $5 \times 10^{51}$ erg ({\it left} panels) and $5 \times
10^{50}$ erg ({\it right} panels). In both cases, the SN outflow
carries more momentum than the beamed ejecta and drives a blast wave
that eventually sweeps up all the GRB-shocked medium.  This happens
before the two beamed blobs collide on the equator. The merged system
will become roughly spherical very soon after. For $E_{\rm SN}= 5
\times 10^{51}\;{\rm erg}$ ($5 \times 10^{50}\;{\rm erg}$), we obtain
$t_{\rm sph} \approx 250\;{\rm yr}$ ($3 \times 10^3\;{\rm yr}$). Not
surprisingly, the presence of an underlying spherical supernova
seriously modifies the simple estimate given by equation (\ref{tsph})
and limits our ability to decipher the presence of a beamed component
in a GRB explosion.
\begin{figure}
\begin{center}
\includegraphics[scale=0.5]{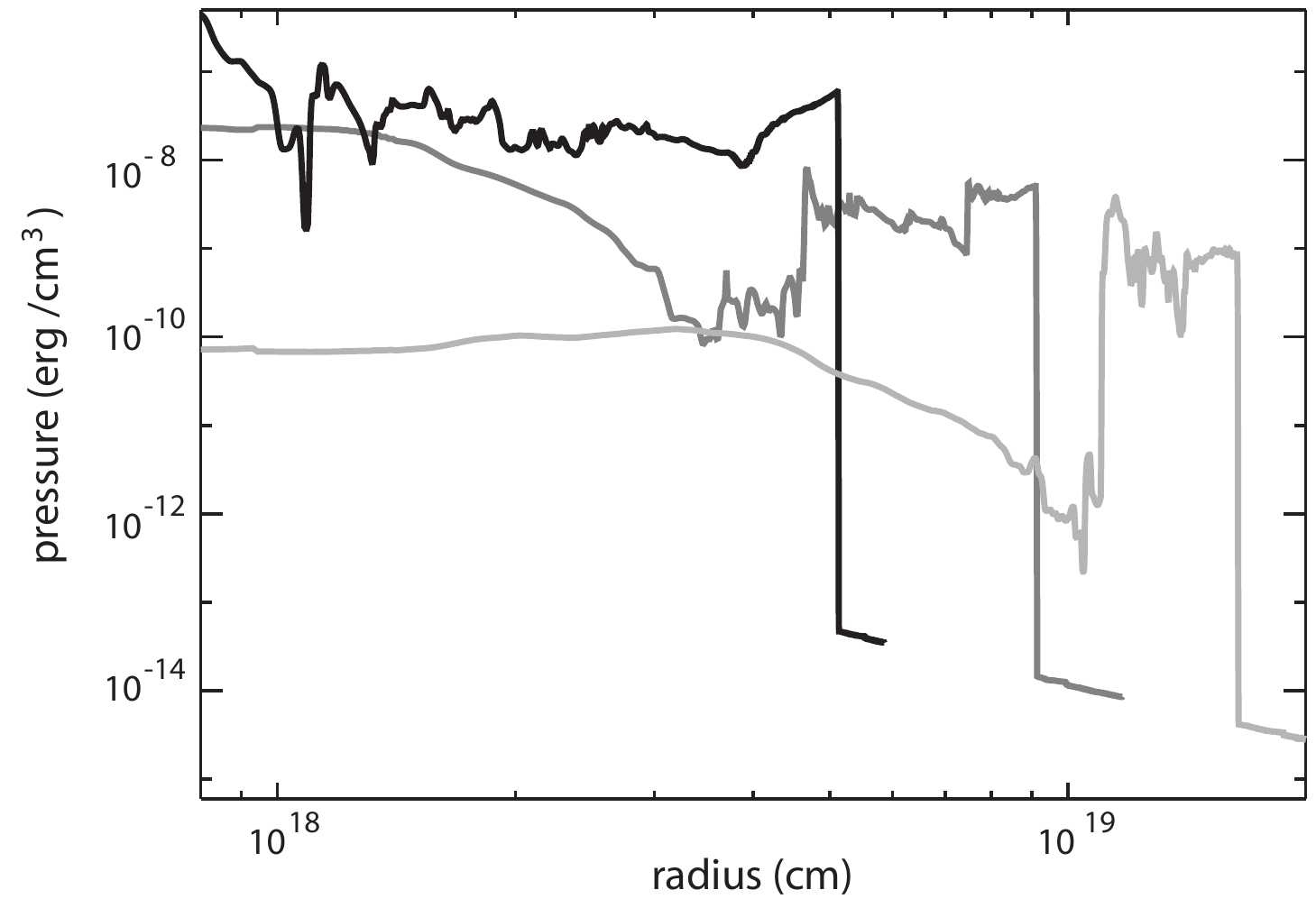}
\caption{The pressure structure, along the $z$ axis of a GRB remnant
interacting with an underlying spherical supernova. The simulation is
the same as in Fig. \ref{fig6} for $E_{\rm SN}= 5 \times 10^{50}$ erg
and $t=51,\;190,\;580$ yr.}
\label{fig7}
\end{center}
\end{figure}

 Prevailing to all these calculations is the initiation of the
 explosion as a pressure-driven blast wave by deposition of the
 explosion energy, $E_{\rm SN}$, entirely as thermal energy. In the
 inner region, an ejecta mass, $M_{\rm SN} \approx 5 M_\odot$, are
 distributed uniformly. This may accurately model the Sedov-Taylor
 stage of SN remnant evolution after the ratio of swept-up mass to the
 mass of the original stellar ejecta exceeds roughly 19
 \citep{fabian}. In most cases, deceleration of the SN outflow will
 begin only sometime after the GRB ejecta has been swept up. As
 clearly seen in Fig. \ref{fig7}, the SN outflow at this early stage
 is not accurately described by the Sedov-Taylor solution. However,
 numerous tests show that our results are not strongly dependent upon
 the assumed mass of the SN ejecta or on whether kinetic energy rather
 than thermal energy is distributed (such that the velocity profile is
 linear; similar to the Sedov solution) in the inner region. None of
 these complications is likely to seriously modified our estimate for
 $t_{\rm sph}$.

\section{GRB$\lowercase{\rm s}$ Inside Pre-Exisiting Wind-Driven Bubbles}\label{bubble}

So far we have considered either the uniform ambient medium case or
the $1/r^2$ wind case on its own. However, since the winds in massive
stars are non-steady, the density structure is more complex
\citep{wij01,ch04,r05,v06}. The preburst stellar wind depends on the
evolutionary stages prior to (and during) the Wolf-Rayet stage
\citep{rr01}. For Galactic stars, a standard evolutionary track is to
start as an O star, evolve through a red supergiant (RSG) phase or
luminous blue variable (LBV) phase with considerable mass loss, and
ending as a Wolf-Rayet star \citep{gs96a,gs96b}. At low metallicity,
the RSG phase may be absent; this may also be the case for some binary
stars.

As an example, we follow the dynamics of a GRB remnant around a 35
$M_\odot$ star \citep[as calculated by][]{ch04}, which evolves (at
solar metallicity) through a long-lived RSG stage with prominent
consequences for the evolution of the circumstellar matter. The wind
velocity in the Wolf-Rayet phase is $10^3$ km s$^{-1}$, and the
mass-loss rate $10^{_5} M_\odot\;{\rm yr}^{-1}$. The ISM pressure and
density are assumed to be typical of the hot, low-density phase of a
starburst galaxy, with $P_{\rm ism}\sim 10^7$ K cm$^{-3}$ and a
density of $4\times 10^{-25}$ g cm$^{-3}$. When the fast Wolf-Rayet
wind $v_w \sim 10^3$ km s$^{-1}$ starts blowing, it sweeps up the RSG
wind material into a shell. The termination shock of the Wolf-Rayet
wind is located at $R_{\rm t}\approx 0.4$ pc and RSG shell at $R_{\rm
rsg}\approx 1.7$ pc. Because the pressure in the shocked wind is
nearly in equilibrium with the ISM, and the temperature $\sim
10^7(v_w/10^3\;{\rm km\;s^{-1}})^2$ K, the density in the bubble is
$\sim 8 \times 10^{25} (P_{\rm ism}/10^7)(v_w/10^3\;{\rm km
s^{-1}})^{-2}\;{\rm g\;cm^{-3}}$, independent of the mass-loss rate
and the ambient density. The extent of the constant density region is
$\sim 4 R_{\rm t}$ out to the dense red supergiant shell.

\begin{figure}
\begin{center}
\includegraphics[scale=0.5]{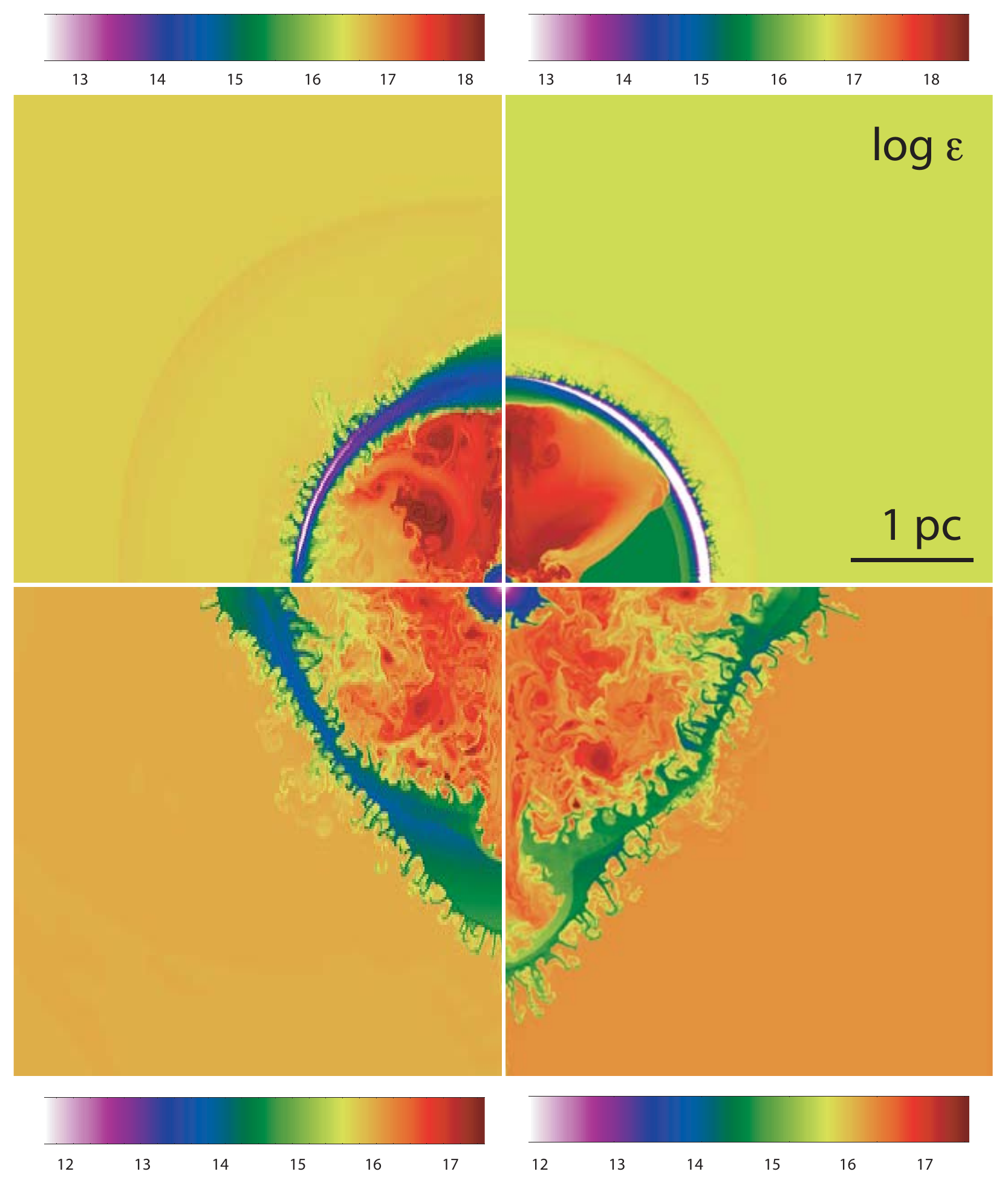}
\caption{The evolution of a GRB remnant inside the wind bubble
structure expected around a 35 $M_\odot$ massive star. The GRB ejecta
initial quantities are the same as in Fig. \ref{fig4}. Shown is the
evolution of the specific energy, $\epsilon$, in erg/g at
$t=51,660,2300$ and 3400 yr. The individual frames have been
successively rotated by $\pi/2$. Calculations were done in
two-dimensional cylindrical coordinates for eight levels of
refinement. The size of the computational domain was (10 pc)$^{2}$. }
\label{fig8}
\end{center}
\end{figure}

\begin{figure}
\begin{center}
\includegraphics[scale=0.55]{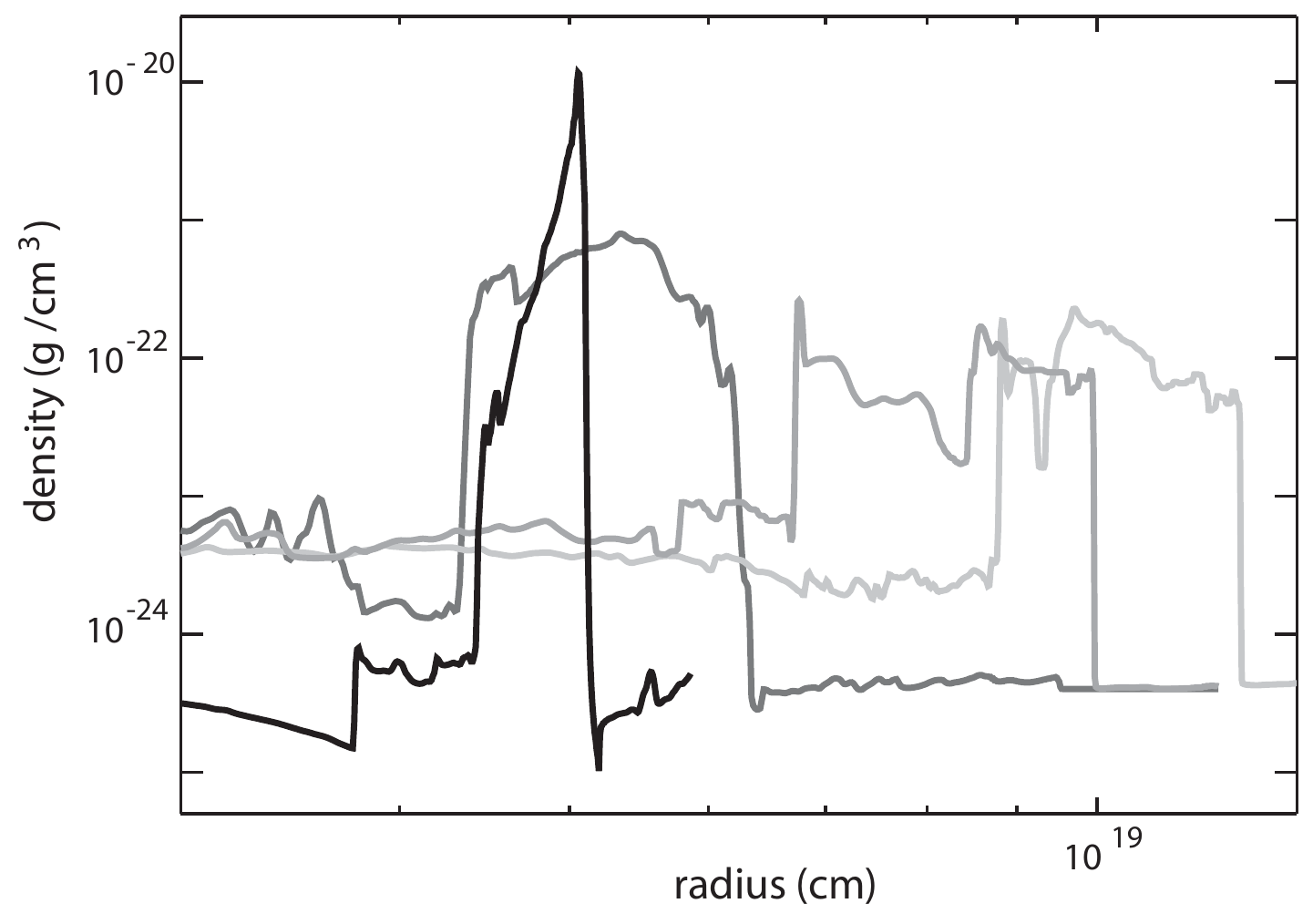}
\caption{The density structure, along the $z$ axis of a GRB remnant
interacting with a wind bubble structure. The simulation is the same as
in Fig. \ref{fig8} for $t=51,660,2300$ and 3400 yr.}
\label{fig9}
\end{center}
\end{figure}
The resulting evolution a beamed GRB remnant in the circumstellar
medium expected around a 35 $M_\odot$ massive stellar progenitor is
summarized in Fig. \ref{fig8}. The presence of the sharp density
gradient will only affect the dynamics of the GRB remnant when it size
is comparable to, or exceeds, the scale length of the gradient. Before
this time, the evolution of the remnant is similar to that depicted in
Fig. \ref{fig4}. The bow shock propagates in the direction of motion
but also perpendicular to it and, over time, wraps around the
expanding ejecta.  In a wind bubble, the shock front will expand
within the stellar wind until it reaches the sharp density
discontinuity at about 1.7 pc. The encounter with the RSG shell
happens before the GRB remnant has time to expand laterally, which
allows for an elongation of the RSG shell in the $z$ axis. A less
pronounced elongation is also seen perpendicular to the $z$ axis,
which results from the collision on the equator of the two beamed
blobs. Superimposed on this large-scale deformation one can also
notice the familiar effects caused by the development of
Rayleigh-Taylor (R-T) instability.  The instability grows rather
quickly and within the following $10^3$ yr several elongated spikes
extend from the shell. At $\approx 2 \times 10^3$ yr, the remnant has
a mean radius of 3 pc (Fig. \ref{fig9}) and the spikes are even more
pronounced than before. Both shell and R-T spikes advance with a
velocity of $300-500$ km s$^{-1}$, which is a very small fraction of
the random velocities observed in the hot cavity of the remnant. These
fast motions are induced by inflection of consecutive shock waves
ramming the irregular shell. The pressure in the cavity is almost
uniform, whereas the density varies chaotically. The knots of ejecta
formed by the R-T instability are believed to be responsible for
several forms of observable radiation in some young SNRs, including
radio synchrotron and optical emission lines. The sheared motions
resulting from the instability could lead to an amplification of the
magnetic field strength and enhance the bright radio remnant
\citep{jn96}.

The growth time of the instability may be roughly estimated in the
following way. The interaction of the GRB remnant with the wind cavity
leads to an increase of the shell-driving pressure by
\begin{equation}
\Delta P={E_{\rm j} \over 2\pi R_{\rm rsg}^3}.
\end{equation}
The mass of the shell is equal to 
\begin{equation}
M_{\rm rsg}={4 \over 3}\pi R_{\rm rsg}^3\rho_{\rm rsg}.
\end{equation}
$\Delta P$ causes an acceleration $g_{\rm rsg}$ of the shell,
satisfying the equation 
\begin{equation}
4\pi R_{\rm rsg}^2 \Delta P=M_{\rm rsg} g_{\rm rsg}.
\end{equation}
With no other forces involved, the contact surface separating both
fluids is unstable to perturbations of all wavelengths and the fluids
interpenetrate. The instability grows exponentially on a
characteristic time scale
\begin{equation}
t_{\rm R-T}= \left({4\pi^2\lambda R_{\rm rsg}^4\rho_{\rm rsg} \over 3
E_{\rm j}}\right)^{1/2},
\label{trt}
\end{equation}
where $\lambda$ is the perturbation wavelength, corresponding to a
growth time of
\begin{equation}
t_{\rm R-T} \approx 10^3\left({R_{\rm rsg} \over 2\;{\rm
pc}}\right)^2\left({\lambda \over 0.2\;{\rm
pc}}\right)^{1/2}\rho_{{\rm rsg},-21}^{1/2}E_{51}^{1/2}\;{\rm yr},
\end{equation}
where $E_{51}$ is the total energy of the GRB remnant and $\rho_{\rm
rsg}=$$10^{-21}\rho_{{\rm rsg},-21}\;{\rm g\;cm^{-3}}$. Indeed, the
observed spikes have dimensions of tenths of parsecs and grow on a
time scale comparable to the estimated one. For this discussion we
have assumed that the remnant evolution is effectively adiabatic and
should be modified to include the effects of radiative
cooling\footnote{For example, the growth of the R-T instability in the
radiating shell is found to be higher than in the adiabatic case
\citep{chb95}.}, which are expected to become important after about
few thousand years.

\begin{figure}
\begin{center}
\includegraphics[scale=0.5]{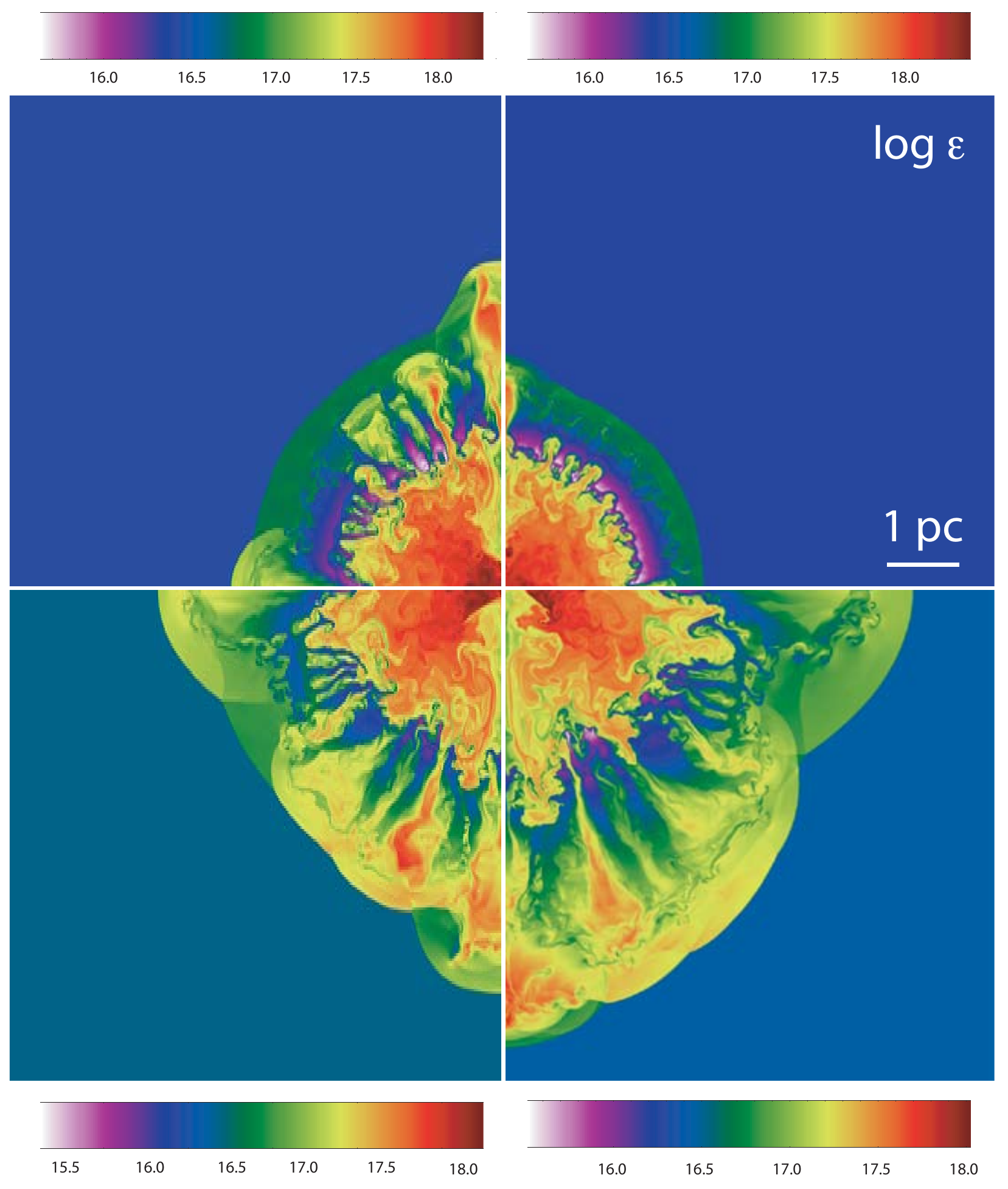}
\caption{The evolution of a GRB remnant accompanied by an underlying
spherical supernova. The wind bubble structure as well as the GRB
ejecta initial quantities are the same as in Fig. \ref{fig8}.  Shown
is the evolution of the specific energy, $\epsilon$, in erg/g at
$t=380,570,765$ and 950 yr. The individual frames have been
successively rotated by $\pi/2$. Calculations were done in
two-dimensional cylindrical coordinates for eight levels of
refinement. The size of the computational domain was (20 pc)$^{2}$. }
\label{fig10}
\end{center}
\end{figure}
\begin{figure}
\begin{center}
\includegraphics[scale=0.55]{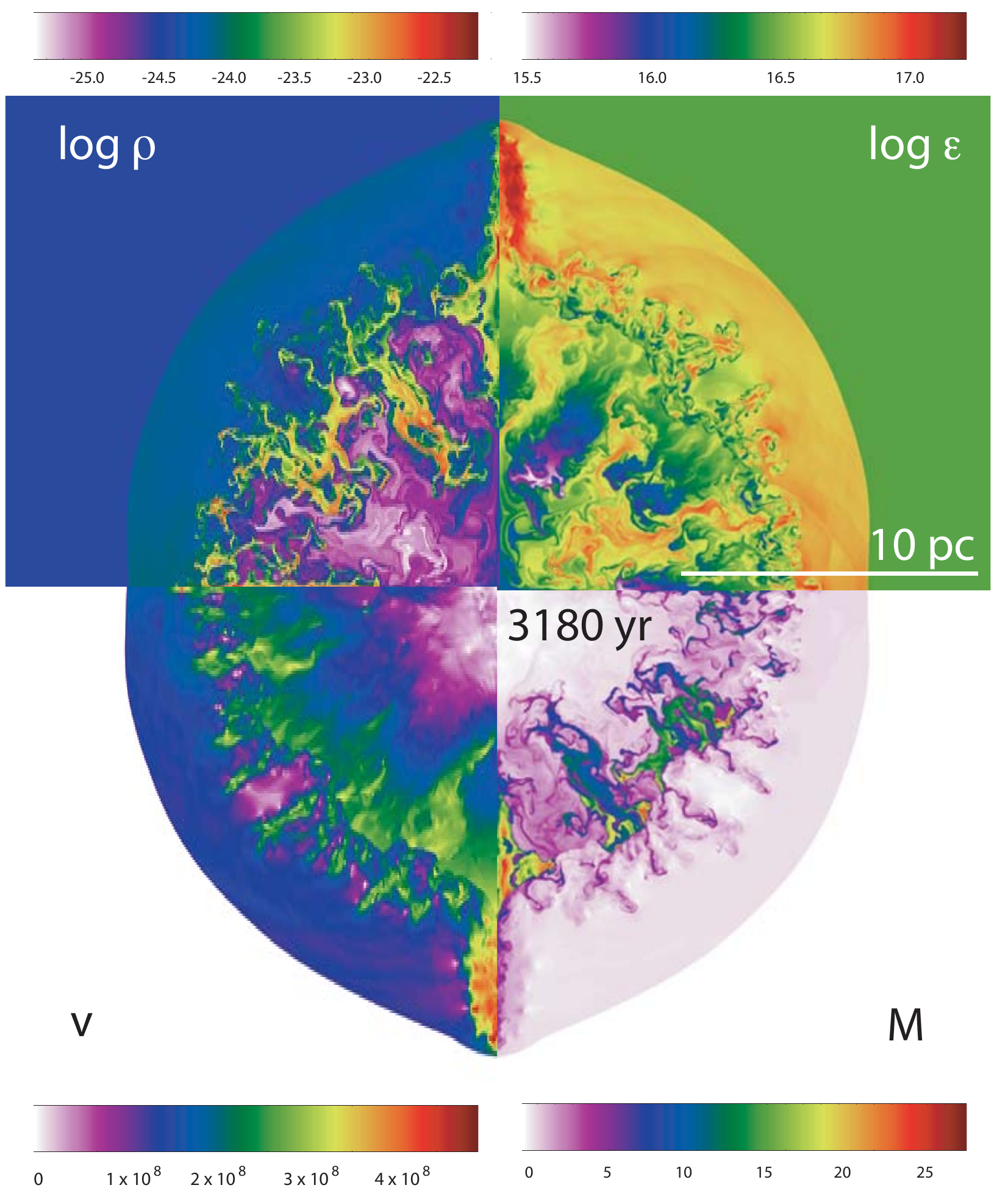}
\caption{The density ($\rho$), specific energy ($\epsilon$), velocity
($v$), and Mach number ($M$) of a GRB remnant accompanied by an
underlying spherical supernova ($M_{\rm SN} \approx 5 M_\odot$)
interacting with a wind bubble structure. The simulation is the same
as in Fig. \ref{fig10} but for $t=3180$ yr.}
\label{fig11}
\end{center}
\end{figure}

The disruption of the RSG shell is, as expected, very sensitive to the
total amount of energy and momentum released by the GRB explosion.  To
illustrate this, we study the evolution of a GRB remnant accompanied
by an underlying $10^{52}$ erg SN in a wind bubble structure such as
that illustrated in Fig. \ref{fig8}. The large-angle SN outflow,
carrying significantly more energy and inertia than the relativistic
jet itself, sweeps up the GRB ejecta before it is decelerated by the
RSG shell. The resultant shell, clearly apparent in the first frame of
Fig. \ref{fig10} taken at $t$=380 yr, will then be pushed outward at
faster velocity than it would be in the absence of the SN ejecta. In
subsequent evolutionary phases one can observe a broadening of the
merged shell, accompanied by a decreased of its density, and the
development of small-amplitude density perturbations. After $t=3180$
yr, the remnant has becomes highly irregular already when it has grown
to $\approx$ 13 pc, and is also rather weakly elongated in the
direction of motion of the beamed ejecta (Fig. \ref{fig11}). There are
also multiple kinematic components within the remnant. Fast-moving
knots and fast-moving flocculi, dense fragments of SN ejecta,
expanding from the explosion center with velocities of several
thousand km s$^{-1}$. Much slower (several hundred km s$^{-1}$)
flocculi are clearly shocked and accelerated remnants of stellar
material ejected by the SN progenitor prior to the explosion. 

The calculations above demonstrate how the measurable properties of
GRB remnants depend on the nature of the progenitor star and the
medium around it. Moreover, counts of GRB remnants as a function of
age may have huge selection effects, as the actual age of the GRB may
be considerably less than the kinematical age estimated from the
radius of the filaments divided by the expansion velocity.  Although
nurture makes a huge difference, hydrodynamical models of young GRB
remnants are also sensitive to the structure of the ejecta, which
might be rather complicated in detail as demonstrated by inspection of
numerical models of SN explosions \citep[e.g.][]{chb95}.

\section{GRB Remnants and Their Detectability}\label{dis}

\subsection{Asymmetric GRB Remnants in the Local Universe}
It is obvious from the discussions in this paper that the dynamics of
GRB remnants are complex, especially because of rich interactions
between the ejecta and the circumburst medium.  The structure of
ejecta is also important, in particular for young GRB
remnants. Abundant confirmation was provided of the important notion
that the morphology and visibility of GRB remnants are determined
largely by their circumstellar environment, from the initial density
gradient created by the progenitor to the effects of large- and
small-scale circumburst structures in later evolution. This
distribution is expected to be nonuniform around massive GRB
progenitors because of the significant mass loss and the dynamical
effects of the stellar winds.

The presence of a density gradient will only affect the dynamics of a
GRB when the remnant size is comparable to, or exceeds, the scale
length of the gradient. Before this time, the density can be treated
as approximately uniform. In the absence of characteristic scales in
stellar ejecta and in the ambient medium, self-similar, spherically
symmetric solutions exist, and they are widely used to interpret
observational data on young GRB remnants. However, even for the
simplest density distributions, we found that the resulting structure
and dynamics of their remnants are very different from the standard
self-similar solutions. This is mainly because at early stages the
morphology of a beamed GRB remnant would be very different from that
of a spherical explosion. In principle, this can be used to identify
those remnants although the dynamical complexity of their sourrounding
circumburst medium seriously limits ourr ability to decipher their
presence, in particular around massive star progenitors.

\begin{deluxetable}{ccc}
\tablewidth{0pc} \tablecaption{Mean space density of asymmetric GRB
remnants.}  \tablehead{\colhead{} & \colhead{$t_{\rm sph}$} &
\colhead{$\phi_{\rm sph}\,[\Re_{\rm grb}=1{\rm Gpc^{-3}\;yr^{-1}}]$}}
\startdata $k=0$ & $3 \times 10^{3}E_{51}^{1/3}\rho_0^{-1/3}$ yr & $3
\times 10^{3}E_{51}^{1/3}\rho_0^{-1/3} f_b^{-1} \Re_{\rm grb}\;{\rm
Gpc^{-3}}$ \\ $k=2$ & $5 \times 10^{4} E_{51}A_\ast^{-1}$ yr & $5
\times 10^{4}E_{51}A_\ast^{-1} f_b^{-1} \Re_{\rm grb}\;{\rm Gpc^{-3}}$
\\ $R_{\rm rsg}\sim$ 1 pc & $10^{2} E_{51}A_\ast^{-1}$ yr & $10^2
E_{51}A_\ast^{-1} f_b^{-1} \Re_{\rm grb}\;{\rm Gpc^{-3}}$\\ $E_{\rm
sn}\sim E_{\rm j,51}$ & $3\times 10^{3}$ yr & $3 \times 10^{3}f_b^{-1}
\Re_{\rm grb}\;{\rm Gpc^{-3}}$ \\ $E_{\rm sn}\geq E_{\rm j,51}$ &
$\leq 10^{3}$ yr & $\leq 10^{3}f_b^{-1} \Re_{\rm grb}\;{\rm
Gpc^{-3}}$\\ \enddata
\end{deluxetable}

The values of the mean space density, $\phi_{\rm sph}$, of asymmetric
GRB remnants expanding into a variety of circumburst environments are
given in Table 1. Taking the long GRB rate expected in the local
universe to be $\Re_{\rm grb}= 0.3$ Gpc$^{-3}$ yr$^{-1}$
\citep{guettal}, an upper limit to the number density of asymmetric
remnants arising from massive stellar progenitors is
\begin{equation}
\phi_{\rm sph} \approx 5 \times 10^{-3}E_{51}A_\ast^{-1}{f_{b}^{-1} \over
100}\;{\rm Mpc^{-3}}.
\end{equation}
This is a generous upper limit, since it assumes that all long GRBs
occur in a free $1/r^2$ stellar wind and that the relativistic
component is energetically dominant. For comparison, an
ultraconservative lower limit to the space density of long GRBs gives
\begin{equation}
\phi_{\rm sph} \approx 10^{-5}{f_b^{-1} \over 100}\;{\rm Mpc^{-3}}.
\label{ratel}
\end{equation}
With the rates given by \citet{guettas}, the frequency of short GRBs
is about 20 times higher than that of long GRBs. The much lower
external density (likely uniform) medium expected around short GRBs
\citep[e.g.,][]{lee05} suggests
\begin{equation}
\phi_{\rm sph} \approx 6 \times 10^{-3}E_{51}^{1/3} \rho_{-3}^{-1/3}
{f_b^{-1} \over 10}\;{\rm Mpc^{-3}},
\end{equation}
where $\rho_{-3}=10^{-3}\rho_0$. Thus the rate of asymmetric short GRB
remnants could be much higher than equation (\ref{ratel}). However,
the remnant's visibility would be highly biased in favor of those with
massive progenitors.\\

\subsection{Observational  Prospects}
The most difficult task at present is to relate hydrodynamical
modeling to observations. A few of the observables, such as expansion
rates and thicknesses of the flow structures, can be relatively easily
determined from the models. However, modeling radio and X-ray emission
is in general difficult, as we are still lacking an understanding of
how electrons are accelerated in shocks. Very similar difficulties are
encountered in modeling nonthermal X-rays. Thermal X-ray spectra are
in principle easier to model, but in practice the difficulties are
formidable. The reason for these difficulties is our poor
understanding of a number of topics, such as the amount of electron
heating in collisionless shocks, the detailed structure and
composition of ejecta, their clumping, the presence of the
inhomogeneous circumstellar medium, and the presence of dust.
\begin{figure}
\begin{center}
\includegraphics[scale=0.55]{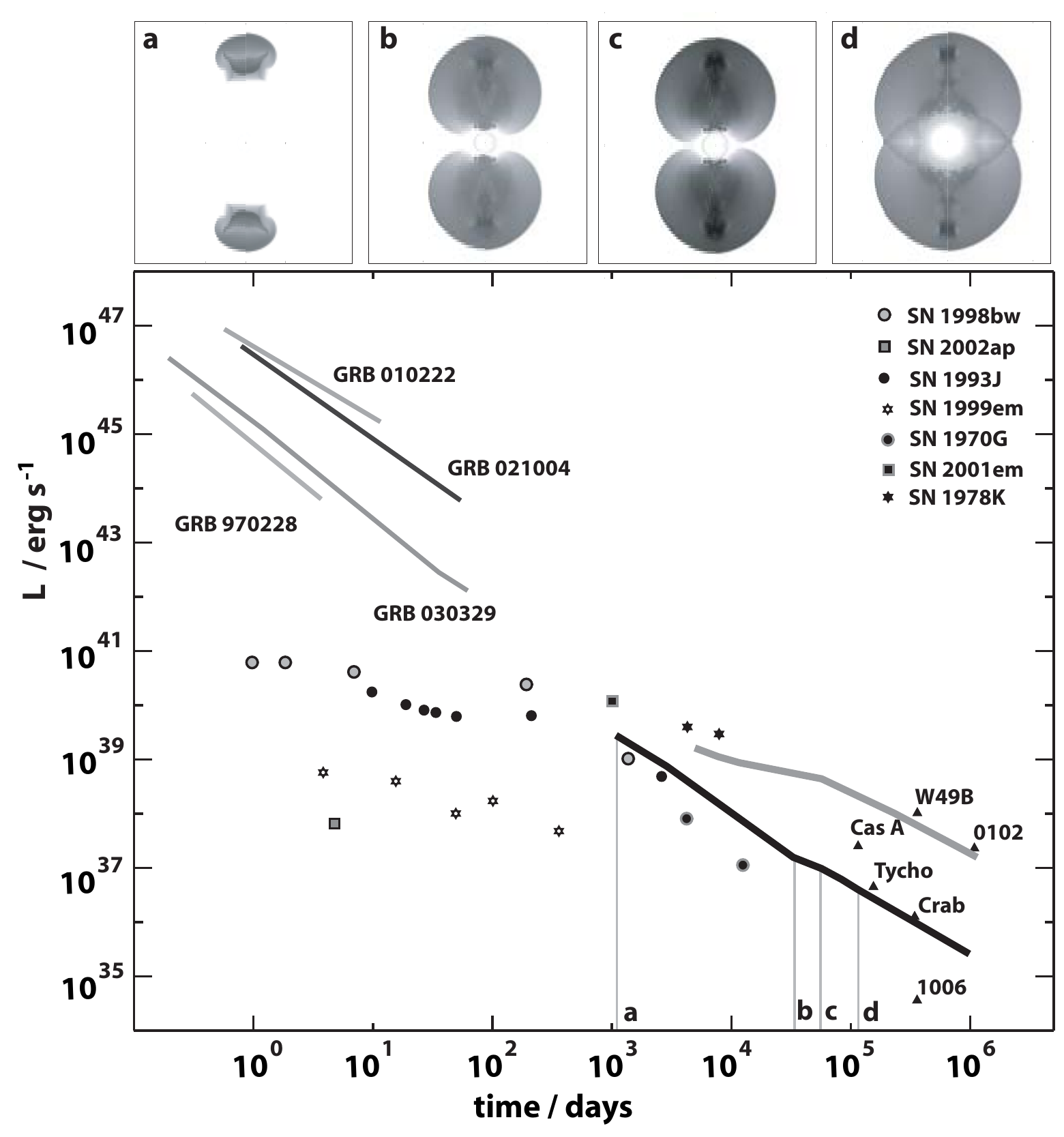}
\caption{ Compilation of GRB, supernova and SN remnant X-ray light
  curves (0.3- 10 keV) presented as (isotropic) luminosity distances
  as a function of age \citep[adapted
    from][]{k04,immler}. Superimposed on these plots, on a common
  scale, is the schematic lightcurves of two GRB remnants expanding
  into different external density environments. The inset figures show
  the evolution of the GRB remnant in a $1/r^2$ density profile (the
  simulation is the same as in Fig. \ref{fig4}). The {\it grey} curve
  shows the evolution of a GRB remnant in a pre-existing wind-bubble
  (the simulation is the same as in Fig. \ref{fig8}). To estimate the
  X-ray lightcurves, we have summed up the internal energy density of
  the gas, $\epsilon_{\rm int}$, for each zone in the simulations and
  plotted $L_X \sim f_X \epsilon_{\rm int}$ as a function of time,
  where $f_X\sim 0.01$ is the fraction of internal energy that is
  radiated away in the 0.3-10 keV range. }
\label{fig12}
\end{center}
\end{figure}

The difficult task of interpreting observations with the help of
hydrodynamical models is perhaps best illustrated by Cas A. This is a
remnant of a massive star explosion and a classic prototype shell
supernova remnant. It has been detected throughout the whole
electromagnetic spectrum \citep{rs48,a80,ah01,f01,h04}.  Observations
in various wavelength bands probe very different components of the
remnant: synchrotron radio emission gives us information about
relativistic electrons, thermal X-ray emission is produced by the bulk
of the shocked hot gas, much cooler gas in radiative shocks emits at
optical wavelengths, and observations in infrared reveal still cooler
gas and dust. However, we still do not understand what is the
relationships between all these features and the remnant's
hydrodynamics \citep[e.g.][]{h04}.

To estimate the emissivity of GRB remnant without undertaking the
complicated effort of calculating X-ray spectra, we have summed up the
internal energy density of the gas, $\epsilon_{\rm int}$, for each
zone in the simulations and plotted $L_X \sim f_X \epsilon_{\rm int}$
as a function of time in Fig. \ref{fig12}, where $f_X\sim 0.01$ is the
fraction of internal energy that is radiated away in the 0.3-10 keV
range. This is only a very approximate procedure and should be taken
as an order of magnitude estimate at present. Superimposed on these
plots, on a common scale, are all GRB afterglows with X-ray luminosity
measurements covering several tens of days. There are, unfortunately,
only a few curves, because such measurements can only be made on GRBs
that are relatively nearby, but their light curves should be
illustrative. We then compared these X-ray lightcurves with those of
supernovae and historical Galactic SN remnants.  The resulting plot is
striking in several ways. Despite the huge disparity in initial
appearance, there are indications of a common convergence of all
classes of phenomena to a common resting place: $L_X \sim
10^{39}-10^{40}$ erg s$^{-1}$ about a few years after the explosive
event. Moreover, it clearly illustrates that the transition from a GRB
to a SN remnant appears to be rather smooth. Clearly, detailed studies
relating hydrodynamical modeling to observations are needed to study
the transition from a GRB into a stellar remnant.

Fig. \ref{fig12} shows the schematic lightcurves of two GRB remnants
expanding into different external density environments. For a $1/r^2$
density profile, as in the {\it black} curve of Fig. \ref{fig12}, the
luminosity of the remnant is initially dominated by the emission of
the individual beamed components as they interact with the stellar
wind. The luminosity continues with a quasi-steady decay rate until
the individual beamed components collide to form a single structure.
The resultant lightcurve will then be characterized by a modest
increase in luminosity. For a GRB expanding inside a pre-existing
wind-bubble, as in the {\it grey} curve of Fig. \ref{fig12}, the
resultant lightcurves can evolve much faster into luminous remnants
such as Cas A \citep{h04} or W49B \citep{mi06} due to their strong
interaction with the dense circumburst medium.

This plot summarized many of the issues outlined in this paper, in
which we argued that the morphology and visibility of GRB remnants are
governed by the pre-existing structure of their {\it birthplace}
environments. Since GRB remnants result from the impact of their
ejecta with circumstellar gas, their visibility is highly biased in
favor of those with massive progenitors. Many young GRBs from massive
progenitors would be bright because their ejected mass is interacting
with nearby gas expelled by the progenitor itself. This circumstellar
gas is likely to have mass comparable to that of the accompanied SN
debris and will not extend much further than a few parsecs. After less
than a century, the blast wave from the GRB will pass through this
relatively dense circumstellar gas. Inferring the the presence of a
beamed component in these hypernova scenarios would be
challenging. The number density of asymmetric GRB remnants in the
local Universe could be far larger if they expand in a tenuous
interstellar medium as expected, for example, in the merger of two
neutron stars (although there are reasons to suspect that the ejecta
may not be too narrowly beamed) and may be easier to constrain
directly (acknowledging the obvious trade-offs in sensitivity and
angular resolution, particularly for radio and X-ray observations).

\acknowledgments We have benefited from many useful discussions with
C. Fryer, J. Granot, T. Piran and S. Woosley.  We are especially
grateful to W. Zhang for countless insightful conversations. The
software used in this work was in part developed by the DOE-supported
ASCI/Alliance Center for Astrophysical Thermonuclear Flashes at the
University of Chicago. Computations were performed on the IAS Scheide
computer cluster. This work is supported by NSF: PHY-0503584 (ER-R)
and DOE SciDAC: DE-FC02-01ER41176 (ER-R and AM).

\end{document}